\documentclass{crmp-l}
\usepackage{latexsym}

\newtheorem{theorem}{Theorem}[section]

\theoremstyle{definition}

\newtheorem{example}[theorem]{Example}

\theoremstyle{remark}

\numberwithin{equation}{section}

\begin{document}

\title{B\"acklund Transformations of Soliton Systems
from Symmetry Constraints}

\author{Wen-Xiu Ma}
\address{
Department of Mathematics, City University of Hong Kong,
Kowloon, Hong Kong, P. R. of China} 
%%\curraddr{Department of Mathematics and Statistics,
%%Case Western Reserve University, Cleveland, Ohio 43403}
\email{mawx@math.cityu.edu.hk}
\thanks{The work was supported by a strategic research grant from City University of Hong Kong (Project No. 7000803)
and a competitive earmarked research grant from the Research Grants Council of Hong Kong 
(Project Nos. 9040395, 9040466)}

\author{Xianguo~Geng}
\address{
Department of Mathematics, Zhengzhou University, Zhengzhou,
P. R. of China}
\email{gengxg@public2.zz.ha.cn}
%%\thanks{Support information for the second author.}

%    General info
%%\date{January 1, 1994 and, in revised form, June 22, 1994.}

%%\dedicatory{This paper is dedicated to our authors.}

\begin{abstract}

Binary symmetry constraints are applied to constructing B\"acklund transformations of soliton systems, both continuous and discrete. Construction of solutions to soliton
systems is split into finding solutions to lower-dimensional Liouville integrable systems, which also paves a way for separation of variables and exhibits 
integrability by quadratures for soliton systems. Illustrative examples are 
provided for the KdV equation, the AKNS system of nonlinear Schr\"odinger equations,
the Toda lattice, and the Langmuir lattice.

\end{abstract}

\maketitle

\section{Introduction}

Symmetry constraints \cite{KonopelchenkoSS-PLA1991,MaS-PLA1994}
play an important role in showing integrability by quadratures for 
soliton systems, both continuous and discrete. They help to generate 
finite-dimensional integrable systems 
\cite{MaS-PLA1994,Cao-SCA1990,CaoG-book1990}
and integrable symplectic mappings \cite{WuG-JMP1996}, and further provide 
a way of constructing finite-gap solutions to soliton systems
by means of Riemann-theta functions \cite{BelokolosBEIM-book1994}.
Based on Lax pairs, symmetries themselves can also be applied to constructing 
B\"acklund transformations of soliton systems \cite{YangS-PLA1994}.
However, there is little work showing the importance of symmetry constraints 
in the study of B\"acklund transformations. 
In this paper we focus on the construction of B\"acklund transformations 
by using symmetry constraints, and show in certain cases that symmetry 
constraints can break up soliton systems into lower-dimensional Liouville integrable systems. 

Let us recall some fundamental concepts. 
A system of continuous equations $u_t=K(u,u_x,\cdots)$ 
is said to have a continuous Lax pair 
\begin{equation}
 \phi _x=U(u,\lambda )\phi,\  \phi_t=V(u,u_x,\cdots;\lambda )\phi , \label{cLaxpair} 
\end{equation} 
if it is equivalent to the compatibility condition $U_t-V_x+[U,V]=0$ of
(\ref{cLaxpair}) under the isospectral condition $\lambda _t=0$,
and a system of 
discrete equations $u_t=K(u,E^{-1}u,Eu,\cdots )$ (where 
$E$ is the shift operator, i.e., $(Eu)(n)=u(n+1)$) is said to have 
a discrete Lax pair 
\begin{equation}
E \phi =U(u,\lambda )\phi,\  \phi_t=V(u,E^{-1}u,Eu,\cdots;\lambda )\phi ,\label{dLaxpair} 
\end{equation}
if it is equivalent to the compatibility condition
$U_t-(EV)U+UV=0$ of
(\ref{dLaxpair}) under the isospectral condition $\lambda _t=0$.
The corresponding adjoint Lax pairs presenting the same compatibility 
conditions read as  
\begin{eqnarray}&& \psi _x=-U^T(u,\lambda )\psi,\  \psi_t=-V^T(u,u_x,\cdots;\lambda
)\psi; \\ &&
 E^{-1}\psi =(E^{-1}U^T(u,\lambda ))\psi,\  \psi_t=-V^T(u,E^{-1}u,Eu,\cdots;\lambda
)\psi; \end{eqnarray}
where $(\cdot)^T$ denotes matrix transpose. 
Adjoint Lax pairs can help us 
determine the variational derivative of the spectral parameter with respect to 
the potential $u$ (see, for example, \cite{FokasA-JMP1982,MaFO-PA1996}
for the continuous case).

A soliton hierarchy of continuous or discrete systems  
\begin{equation}
u_{t_n}=K_n=\Phi ^nK_0=JG_n=J\frac {\delta {\tilde H}_n}{\delta u},\ n\ge 0,
\end{equation}
can be generated through the isospectral ($\lambda _{t_n}=0$) compatability conditions of continuous Lax pairs  
\begin{equation}
 \phi _x=U(u,\lambda )\phi,\  \phi _{t_n}=V^{(n)} (u,u_x,\cdots;\lambda 
)\phi,\end{equation}
or discrete Lax pairs  
\begin{equation}
 E\phi =U(u,\lambda )\phi,\  \phi _{t_n}=V^{(n)} (u,E^{-1}u,Eu,\cdots;\lambda 
)\phi, \end{equation}
where $V^{(n)}$ are Laurent polynomials in $\lambda $,
$\Phi $ is a hereditary recursion operator to map symmetries into symmetries,  
and $J$ is a Hamiltonian operator to map conserved covariants to symmetries.

In this paper, we would like to show that symmetry constraints can be applied to 
constructing B\"acklund transformations of soliton systems from Lax pairs.
The resulting B\"acklund transformations separate each soliton system 
(in a hierarchy) 
into two lower-dimensional integrable systems. Thus, symmetry constraints 
are shown to be very useful in exposing integrability by quadratures 
for soliton systems. Illustrative examples will be given in both continuous and 
discrete cases. 

\section{Symmetry constraints}

Let us consider the space parts and the time parts
of Lax pairs and adjoint Lax pairs: 
\[ \left\{\begin{array} {l} \phi _x =U(u,\lambda )\phi , \vspace{2mm}\\ \psi
_x=-U^T(u,\lambda )\psi ;\end{array} \right.  \quad \left\{
 \begin{array} {l} \phi _{t_n} =V^{(n)}(u,u_x,\cdots;\lambda
)\phi , \vspace{2mm}\\ \psi _{t_n}=-V^{(n)T}(u,u_x,\cdots;\lambda )\psi ;\end{array} \right.  \]
or 
\[   \left\{\begin{array} {l} E\phi  =U(u,\lambda )\phi , \vspace{2mm}\\ E^{-1}
\psi =(E^{-1}U^T(u,\lambda ))\psi ;\end{array} \right.  \quad \left\{
 \begin{array} {l} \phi _{t_n} = V^{(n)}(u,E^{-1}u,Eu,\cdots;\lambda
)\phi , \vspace{2mm}\\ \psi _{t_n}=-V^{(n)T}(u,E^{-1}u,Eu,\cdots;\lambda )\psi .\end{array} \right.  \]
By using the space parts, we can work out 
\begin{equation}
 \frac {\delta \lambda }{\delta u}=\alpha   \psi ^T\frac {\partial
U(u,\lambda )}{\partial u} \phi  ,
\ \textrm{or}\ \frac {\delta \lambda }{\delta u}=\beta   (E\psi ^T)
\frac {\partial U(u,\lambda )}{\partial u} \phi  ,\label{deltalambda}
\end{equation}
where $\alpha $ and $\beta$ are two constants.
Note that the Lie homomorphism $J\frac {\delta }{\delta u}$
transforms conserved functionals to symmetries.
Therefore, $J\frac {\delta \lambda }{\delta u} $ is a symmetry of each system
$u_{t_n}=K_n$, since $\lambda $ is a conserved functional, i.e.,
$\lambda _{t_n}=(\lambda (u))_{t_n}=0$ when $u_{t_n}=K_n$.

Now for all $m_0\ge 0$, we can make symmetry constraints
\begin{equation}
 K_{m_0}= \frac 1 \alpha 
J\frac {\delta \lambda }{\delta u} = J\psi ^T\frac {\partial
U(u,\lambda )}{\partial u}\phi ,\ \textrm{or} \ 
K_{m_0}= \frac 1 \beta 
J\frac {\delta \lambda }{\delta u} = J(E\psi ^T)\frac {\partial
U(u,\lambda )}{\partial u}\phi .\end{equation}
If we take distinct eigenvalues
$\lambda _1,\cdots, \lambda _N$, and suppose that  
\begin{equation} \phi^{(j)} _{x}=U(u,\lambda _j) \phi
^{(j)},\ \psi ^{(j)}_{x}=-U^T(u,\lambda _j) \psi^{(j)}, \end{equation}
or 
\begin{equation} E\phi^{(j)} =U(u,\lambda _j) \phi
^{(j)},\ E^{-1}\psi ^{(j)}=(E^{-1}U^T(u,\lambda _j)) \psi^{(j)}, \end{equation}
where \[
\phi^{(j)}=(\phi_{1j},\cdots,\phi_{sj})^T, \ 
\psi^{(j)}=(\psi_{1j},\cdots,\psi_{sj})^T,\]
we can make more systematical symmetry constraints
\begin{equation} K_{m_0} = \sum_{j=1}^N \frac 1 {\alpha _j} 
J \frac {\delta \lambda _j}{\delta u},\ \textrm{or}\ 
K_{m_0} = \sum_{j=1}^N \frac 1 {\beta _j} 
J \frac {\delta \lambda _j}{\delta u},
\nonumber \end{equation} 
namely, 
\begin{equation}
K_{m_0} = \sum_{j=1}^N 
J  \psi^{(j)T} \frac {\partial  U(u,\lambda _j)}{\partial u}\phi^{(j)},
\ \textrm{or}\ K_{m_0} = \sum_{j=1}^N J
 (E\psi^{(j)T}) \frac {\partial U(u,\lambda _j)}{\partial u}\phi^{(j)},
\label{symmetryconstraints}
\end{equation}
where $\alpha _j$ and $\beta _j$ are the constants defined as in 
(\ref{deltalambda}). These symmetry constraints suggest 
\begin{equation}
G_{m_0}=\sum_{j=1}^N \psi
^{(j)T}\frac {\partial U(u,\lambda _j)}{\partial u}\phi^{(j)},
\ \textrm{or}\ 
G_{m_0} =\sum_{j=1}^N (E\psi
^{(j)T})\frac {\partial U(u,\lambda _j)}{\partial u}\phi^{(j)}
. \label{sc} \end{equation}
Among those constraints between the potential, $u$, and
the eigenfunctions and adjoint eigenfunctions, $\phi^{(j)}$ and
$\psi^{(j)}$,
the Bargmann constraint 
will be applied to constructing B\"acklund 
transformations between soliton systems and lower-dimensional integrable systems.

\section{B\"acklund transformations}

Let us take the Bargmann constraint, i.e., the constraint
(\ref{sc}) with $ G_{m_0} =G_{m_0}(u)$ not involving any $\partial ^i u$
($\partial =\frac \partial {\partial x}$), $i>0$, or any $E^iu$, $i\ne 0$. 
Note that the discrete constraint defined by (\ref{sc}) 
can be rewritten as 
\begin{equation}
G_{m_0}(u)=\frac{\delta {\tilde H}_{m_0}(u)}{\delta u}
=\sum_{j=1}^N \frac 1 {\beta _j} \frac {\delta \lambda _j}{\delta u}
 =\sum_{j=1}^N 
\psi^{(j)T}U^{-1}(u,\lambda _j)\frac {\partial U(u,\lambda _j)}{\partial u}\phi^{(j)},
\end{equation}
and therefore, in each of continuous and discrete cases,
the Bargmann constraint defined by (\ref{sc}) is an algebraic equation
on $u$, $\phi^{(j)}$, and $ \psi ^{(j)}$. Assume that solving the corresponding algebraic 
equation for $u$ gives rise to an explicit expression of $u$:
\begin{equation}
 u=f (\phi ^{(1)},\phi
^{(2)},\cdots,\phi ^{(N)};\psi^{(1)},\psi^{(2)}, \cdots,\psi^{(N)}).
\label{BT} 
\end{equation}
Substituting this expression of $u$ into Lax pairs and adjoint Lax pairs
leads to two systems, called binary constrained Lax pairs.
Binary constrained continuous Lax pairs read as
\begin{eqnarray}&&  
\left \{ \begin{array}{l}
\phi^{(j)} _{x}=U(f ,\lambda _j) \phi ^{(j)},\ 1\le j\le N,\vspace{2mm}\\ \psi
^{(j)}_{x}=-U^T(f,\lambda _j) \psi^{(j)},\ 1\le j\le N;
\end{array}\right. \label{xpartofbclp} \\ &&
 \left
\{\begin{array}{l} \phi ^{(j)}_{t_n}=V^{(n)}(f ,f_x,\cdots;\lambda _j) \phi ^{(j)} 
,\ 1\le j\le N,\vspace{2mm}\\ \psi
^{(j)}_{t_n}=-V^{(n)T}(f ,f_x,\cdots;\lambda _j) \psi ^{(j)},\ 1\le j\le N;\end{array}\right.
\label{tpartofbclp}   \end{eqnarray}
the first of which is a system of ordinary differential equations, 
but the second of which is a system of partial differential equations
since $V^{(n)}$ contains some derivatives of u with respect to $x$.
Binary constrained discrete Lax pairs read as
\begin{eqnarray} &&
\left \{ \begin{array}{l}
E\phi^{(j)} =U(f ,\lambda _j) \phi ^{(j)},\ 1\le j\le N,\vspace{2mm}\\ E^{-1}\psi
^{(j)}=(E^{-1}U^T(f ,\lambda _j)) \psi^{(j)},\ 1\le j\le N;
\end{array}\right.
\label{Epartofbclp}
\\ &&\left
\{\begin{array}{l} \phi ^{(j)}_{t_n}=V^{(n)}(f ,E^{-1}f,Ef,\cdots;\lambda _j)  \phi ^{(j)} 
,\ 1\le j\le N,\vspace{2mm}\\ \psi
^{(j)}_{t_n}=-V^{(n)T}(f ,E^{-1}f,Ef,\cdots;\lambda _j) \psi ^{(j)},\ 1\le j\le N;\end{array}\right.
\label{tEpartofbclp}  \end{eqnarray} 
the first of which is a system of difference equations, 
but the second of which is a system of difference-differential equations since $V^{(n)}$ contains some of $E^iu$, $i\ne 0$.
However, the second systems can be tranformed into systems 
of ordinary differential equations by using the first systems.
Furthermore, it can be shown by $r$-matrix formulation
that all binary constrained Lax pairs,
both continuous and discrete, are integrable in the Liouville sense \cite{Arnold-book1989,BruschiRST-PD1991}. 

Therefore, (\ref{BT}) provides 
B\"acklund transformations between soliton systems 
and integrable binary constrained Lax pairs,
and construction of solutions $u=f(\phi^{(j)},\psi^{(j)})$ to soliton systems
is split into finding solutions $\phi^{(j)}$ and $\psi^{(j)}$ to two lower-dimensional
integrable systems.

\section{Examples of continuous systems}

\begin{example} Let us consider the KdV Equation 
\begin{equation}
u_{t_1}=\frac 14 u_{xxx}+\frac 32 uu_x= J\frac {\delta {\tilde H}_1}{\delta u},\
J=\partial , \ {\tilde H}_1=\int (\frac 18 uu_{xx}+\frac 3{12} u^3) \,dx,
\label{KdV}\end{equation}
which can be written as $U_t-V_x+[U,V]=0$ with 
\begin{equation} 
U=\left(\begin{array} {cc}0&1\vspace{2mm} \\ \lambda -u & 0 \end{array} \right)
,\ V= \left(\begin{array} {cc}
-\frac 14 u_x&\lambda +\frac12 u\vspace{2mm} \\ \lambda^2 -\frac 12
u\lambda -\frac 14 u_{xx}-\frac 12 u^2 & \frac 14 u_x \end{array} \right) 
. \label{UVofKdV}\end{equation} 
Take the Bargmann symmetry constraint
\begin{equation}
K_{m_0}= 2\partial _x \sum_{j=1}^N
\psi ^{(j)T}\frac {\partial U(u,\lambda _j)}{\partial u}\phi ^{(j)} 
,\ \textrm{where}\ K_{m_0}=K_0=u_x,\nonumber 
\end{equation}
which implies the following equation
\[ u_x= 
2\partial _x \sum_{j=1}^N
(\psi_{1j} ,\psi _{2j}) 
\left(\begin{array} {cc}0& 0 \vspace{2mm} \\ -1 & 0 \end{array} \right)
\left(\begin{array} {c} \phi_{1j} \vspace{2mm} \\ \phi_{2j} \end{array} 
\right)=2\partial _x \sum_{j=1}^N  \phi_{1j}\psi_{2j} .\]
Integrating this equation with respect to $x$ yields a B\"acklund transformation
\begin{equation}
u=f(\phi _{ij};\psi _{ij})
=2\sum_{j=1}^N \phi_{1j}\psi_{2j}=2<P_1,Q_2>,
\label{BTofKdV}
\end{equation}
where $<\cdot,\cdot>$ denotes the standard inner product of 
${\mbox{\rm I \hspace{-0.9em} R}^N}$, and $P_i$ and 
$Q_i$ are defined by 
\[ P _i=(\phi _{i1},\phi _{i2},\cdots,\phi
_{iN})^T,\ Q_i=(\psi_{i1},\psi_{i2},\cdots,\psi_{iN})^T ,\ i=1,2 .\]
A general B\"acklund transformation $u=2<P_1,Q_2>+c$ with an arbitrary constant $c$ can 
also be resulted from the above symmetry constraint, but it will not generate 
essentially new integrable systems from Lax pair of the KdV equation
and so we omit to discuss it.

Keeping (\ref{BTofKdV}) in mind, 
the corresponding constrained Lax pairs (\ref{xpartofbclp}) and 
(\ref{tpartofbclp}), where two matrices $U$ and $V^{(n)}=V$ are defined by (\ref{UVofKdV}), 
can simultaneously be rewritten as
 \begin{eqnarray}  &&  
P _{ix}=-\frac {\partial H^c}{\partial Q _{i}},\ 
Q _{ix}=\frac {\partial 
H^c}{\partial P _{i}},\ H^c=-F_3,
\ i=1,2, \label{H(x)} \\ &&
P _{it_1}=-\frac {\partial H_1^c}{\partial Q _{i}},\ 
Q _{it_1}=\frac {\partial H_1^c}{\partial P
_{i}},\ H_1^c=-F_4,\ i=1,2 , \label{H_1(t)} \end{eqnarray} 
%with the Hamiltonian functions
%\begin{eqnarray}  && H=-<P _2,Q_1>+<P _1,Q_2>^2-<AP _1,Q_2>, \nonumber \\ &&
%H_1=-\frac 14 <P _1,Q_1>^2-\,\frac 14 <P _2,Q_2>^2 \nonumber \\ &&
%\qquad +\frac 12 <P _1,Q_1><P _2,Q_2> -<P _1,Q_2><P _2,Q_1>
%\nonumber \\ && \qquad -<AP _2,Q_1>-<A^2P _1,Q_2>+<P _1,Q_2> <AP
%_1,Q_2> .  \end{eqnarray}  
where the functions $F_m$ are given by 
\begin{eqnarray}&& 
%F_m=\sum _{i=0}^m(a_ia_{m-i}+b_ic_{m-i}),\vspace{2mm}\\
 F_0=0,\  F_1=1,
\  F_2=0, \nonumber \\ &&  
 F_3= <P _2,Q _1>-<P _1,Q _2>^2+<AP _1,Q  _2>,\nonumber 
\\ &&F_m=
\sum_{i=0}^{m-4}( \bar {a}_i\bar {a}_{m-i-4}+\bar {b}_i\bar {c}_{m-i-4})+\bar {c}_{m-3}+\bar {b}_{m-2}- \frac 12 f \bar {b}_{m-3},\ m\ge 4,  \nonumber 
\\ &&
\bar {a}_i=<A^{i}P _1,Q _1>-<A^{i}P _2,Q _2>,\
\bar {b}_i= <A^{i}P _1,Q _2>,\ \bar {c}_i= <A^{i}P _2,Q _1>,
\nonumber \end{eqnarray}
through $F=\textrm{det}V|_{u=f}=\sum_{m=0}^\infty 
F_m\lambda ^{-m},\ V=\sum_{i=0}^\infty V_i\lambda ^{-i}$ satisfying $V_x=[U,V]$.
Note that we always accept 
\[
A=\textrm{diag}(\lambda _1,\cdots,\lambda _N).\]
These two systems (\ref{H(x)}) and (\ref{H_1(t)}) are Liouville
integrable \cite{Arnold-book1989}, since they have involutive integrals of motion
$F_m, \ m\ge 0,$ and 
\begin{equation}
\bar {F}_j =\phi_{1j}\psi_{1j}+\phi_{2j}\psi_{2j},\ 1\le j\le N, \label{barF_m}
\end{equation}
among which $F_3,F_4,\cdots, F_{N+2}$ and $\bar {F}_1,\cdots ,\bar{F}_N$ 
are functionally independent. 
%The Hamiltonian function $H=-F_3, H_1=\cdots$. 
%The whole process above is called binary nonlinearization\cite{Ma.
All functions 
\begin{equation} u(x,t_1)=2<P _1(x,t_1),Q_2(x,t_1)> 
 =<g_H^xg_{H_1}^{t_1}P _{10}, g_H^xg_{H_1}^{t_1}Q _{20}> \end{equation}
where $g^x_H$ and $g_{H_1}^{t_1}$ are the Hamiltonian flows associated with 
(\ref{H(x)}) and (\ref{H_1(t)}) respectively,
and $P_{10}$ and $Q_{20}$ are two arbitrary constant
vectors, will determine 
solutions to the KdV equation (\ref{KdV}).

\end{example}

\begin{example}
Let us now consider the coupled nonlinear Schr\"odinger system: 
\begin{equation}
u_{t_2}=\left(\begin{array} {c}q\vspace{2mm}\\ r\end{array} \right)_{t_2}= \left ( \begin{array} {c} -\frac12
q_{xx}+q^2r\vspace{2mm} \\ \frac12 r_{xx}-qr^2 \end{array} \right)
=J\frac {\delta {\tilde H}_2}{\delta u}  \label{cnss}\end{equation}
with the Hamiltonian operator $J$ and the Hamiltonian functional
${\tilde H}_2$:
\begin{equation}
 J= \left(\begin{array} {cc}0&-2\vspace{2mm}\\ 
2&0 \end{array} \right),\ {\tilde H}_2=\frac 1 {12}\int (qr_{xx}-q_xr_x+q_{xx} r -3q^2r^2)\,dx.
\end{equation}
It has a Lax pair 
\begin{equation}
%\left(\begin{array}{c} \phi _1 \\ \phi _2 \end{array}\right) _x
%=U\left(\begin{array}{c} \phi _1 \\ \phi _2 \\ \phi_3\end{array}\right)
U=\left(\begin{array}{cc} -\lambda &\ q \vspace{1mm}\\ r & \lambda
\end{array}\right),
%\left(\begin{array}{c} \phi _1 \\ \phi _2 \end{array}\right),
%\ \left(\begin{array}{c} \phi _1 \\ \phi _2 \end{array}\right) _{t_2}
%=V^{(2)}\left(\begin{array}{c}\phi_1 \\ \phi_2\\ \phi_3\end{array}\right)
\ V=\left(\begin{array}{cc} -\lambda ^2+\frac 12 qr & \ q\lambda -\frac{1}{2}q_x
\vspace{2mm}\\ r \lambda +\frac{1}{2}r_x & \ \lambda ^2-\frac 1 2 qr
\end{array}\right).
%\left(\begin{array}{c} \phi _1 \\ \phi _2 \end{array}\right).
\label{UVofcnss}\end{equation} 
The Bargmann symmetry constraint (\ref{symmetryconstraints}) reads as
\begin{equation}  K_{m_0}
%=J\frac {\delta H_0 }{\delta u} %=J\sum _{j=1}^N E_j \frac {\delta \lambda _j}{\delta u}
=J\sum_{j=1}^N
\psi ^{(j)T}\frac {\partial U(u,\lambda _j)}{\partial u}\phi ^{(j)} 
,\ \textrm{where}\ K_{m_0}=K_0
=J\left (\begin{array} {c} r\\ q \end{array} \right).\nonumber 
\end{equation}
This implies the following equation
\begin{equation}
J\left (\begin{array} {c} r\\ q \end{array} \right)=
J \sum _{j=1}^N 
\left (\begin{array} {c} 
(\psi_{1j} ,\psi _{2j}) 
\left(\begin{array} {cc}0& 1 \vspace{2mm} \\ 0 & 0 \end{array} \right)
\left(\begin{array} {c} \phi_{1j} \vspace{2mm} \\ \phi_{2j} \end{array}
\right)\vspace{2mm}
\\
(\psi_{1j} , \psi _{2j}) 
\left(\begin{array} {cc}0& 0 \vspace{2mm} \\ 1 & 0 \end{array} \right)
\left(\begin{array} {c} \phi_{1j} \vspace{2mm} \\ \phi_{2j} \end{array}
\right)
\end{array} \right)
= J\sum _{j=1}^N \left(\begin{array}{c} \phi
_{2j}\psi _{1j}\\ \phi _{1j}\psi_{2j}\end{array}\right),
\nonumber \end{equation}
which equivalently engenders a B\"acklund transformation 
\begin{equation} u =\left (\begin{array} {c} q\\ r \end{array} \right)
=f(\phi_{ij};\psi_{ij})=
\sum_{j=1}^N \left(\begin{array}{c} \phi
_{1j}\psi _{2j}\\ \phi _{2j}\psi_{1j}\end{array}\right)=
\left(\begin{array}{c} < P _1,Q_2>\\ < P _2,Q _1>
\end{array}\right). \label{BTcnss} \end{equation} 

Keeping (\ref{BTcnss}) in mind, 
the corresponding constrained Lax pairs 
(\ref{xpartofbclp}) and 
(\ref{tpartofbclp}), where two matrices $U$ and $V^{(n)}=V$ are defined by (\ref{UVofcnss}), 
can simultaneously be transformed into the following
\begin{eqnarray}  &&  P_{ix}=-\frac {\partial H^c}{\partial Q_i} , \
Q_{ix}=\frac {\partial H^c}{\partial P_i} ,\ H^c=F_2-\frac 14 F_1^2 ,\ i=1,2,  
\label{xpartofclp2}\\&& 
P_{it_2}=-\frac {\partial H_{2}^c}{\partial Q_i} , \ Q_{it_2}=\frac {\partial 
H_{2}^c}{\partial P_i} ,\ H^c_{2}= F_3- \frac 12 F_1F_2+\frac
3{24}F_1^3,\ i=1,2,\label{tpartofclp2}
\end{eqnarray}  
where the functions $F_m$ are given by
\begin{eqnarray}  
&& F_1=<P_2,Q_2>-<P_1,Q_1>,\nonumber 
\\ && F_m=\sum_{i=0}^{m-2}\Bigl[\Bigr.\frac 14 (<A^{i}P_1,Q_1>-<A^{i}P_2,Q_2>)(<A^{m-i-2}P_1,Q_1>  
\nonumber  \\ && \qquad 
-<A^{m-i-2}P_2,Q_2>) +<A^{i}P_1,Q_2><A^{m-i-2}P_2,Q_1>\Bigl. \Bigl] 
\nonumber  \\ && \qquad 
 + <A^{m-1}P_2,Q_2>-<A^{m-1}P_1,Q_1>,\ m\ge 2. \nonumber 
\end{eqnarray}  
These two systems are completely integrable in the Liouville sense 
\cite{Arnold-book1989}, since they have involutive integrals of motion
$F_m$, $m\ge 0$, defined above,
and $\bar{F}_j,\ 1\le j\le N,$ defined by (\ref{barF_m}),
among which $F_1,F_2,\cdots, F_{N}$ and $\bar {F}_1,\cdots ,\bar {F}_N$  
are functionally independent. The B\"acklund transformation (\ref{BTcnss})
determines solutions to the coupled nonlinear Schr\"odinger system (\ref{cnss}):
\begin{equation}\left\{\begin{array}{l}
q(x,t_2)=<P_1(x,t_2),Q_2(x,t_2)>=
\displaystyle{
\sum_{j=1}^N}\phi_{1j}(x,t_2)\psi_{2j}(x,t_2),
\vspace{2mm}\\ 
r(x,t_2)=<P_2(x,t_2),Q_1(x,t_2)>= 
\displaystyle{\sum_{j=1}^N}
\phi_{2j}(x,t_2)\psi_{1j}(x,t_2),
\end{array}\right.
\end{equation}
if $\phi_{ij}(x,t_2)$ and $\psi_{ij}(x,t_2)$ 
simultaneously
solve two integrable finite-dimensional 
Hamiltonian systems (\ref{xpartofclp2}) and (\ref{tpartofclp2}).
\end{example}

\section{Examples of discrete systems}

\begin{example} Let us consider the Toda lattice \cite{Toda-book1989}:
\begin{equation}
a_t(n,t)=a(n,t)(b(n+1,t)-b(n,t)),\quad b_t(n,t)=a(n,t)-a(n-1,t),
\label{(20)}
\end{equation}
which associates with the discrete spectral problem   
\begin{equation}
E{\phi}=U(u,\lambda){\phi},\enspace  U(u,\lambda)=\left(\begin{array} {cc} 0&a\vspace{2mm}
\\ -1&\lambda-b\end{array} \right),\enspace  {\phi}=\left(\begin{array} {c} \phi_{1}\\ {\phi_{2}}\end{array} \right), \label{(1)}
\end{equation}  
where $u=(a,b)^T$ and ${\lambda}$ is a spectral parameter. In order to derive a hierarchy of lattice equations associated with (\ref{(1)}), 
we first solve the stationary discrete zero-curvature equation: 
\begin{equation}
(EV)U-UV=0,\enspace V=(V_{ij})_{2\times 2},\label{(2)}\end{equation}
by assuming 
$$ V_{11}=aB+(b-\lambda)C,\quad V_{12}=E^{-1}aC,\quad V_{21}=-C,\quad V_{22}=E^{-1}aB ,$$
where 
$$ B=\sum_{i\geq 0}B_{i-1}\lambda^{-i},\quad C=\sum_{i\geq 0}C_{i-1}\lambda^{-i}.$$
The discrete zero-curvature equation (\ref{(2)}) requires 
\begin{equation}
JG_{-1}=0,\enspace MG_{n-1}=JG_{n},\enspace  n\geq 0.\label{(3)}\end{equation}
where $G_n=(B_n,C_n)^T$, and $J$, $M$ are two skew-symmetric operators:
\begin{equation}
J=\left(\begin{array} {cc} 0 & a\Delta \vspace{2mm}\\
 -\Delta^*a & 0 \end{array} \right),\enspace 
M=\left(\begin{array} {cc}
a(\Delta-\Delta^*)a&a\Delta b\vspace{2mm}\\ 
-b\Delta^*a&a\Delta-\Delta^*a \end{array} \right)
.\end{equation} 
We choose $G_{-1}=(0,1)^T$, and assume that all terms of $G_n$, $n\ge 0$,
do not belong to $\rm{ker}J=\textrm{span}\{G_{-1},G_{-2}\}$ where 
${G_{-2}=(a^{-1},0)^T,}$
%(\ref{(3)}) has an additional condition $\{G_j\}=\{G_j\}\backslash\{{\rm ker}J\}$. 
when we uniquely determine all $G_n$, $n\ge 0$. 
For instance, the second member has to be $G_0=(1,b)^T.$
This requirement also means that we just choose the key vector fields to form 
systems of lattice equations.

Let $\lambda_1,\cdots,\lambda_N$ be 
distinct eigenvalues. Then we have 
\begin{eqnarray}&&
(E\phi_{1j},E\phi_{2j})=(\phi_{1j},\phi_{2j})U(u,\lambda_{j})^{T},\enspace 1\leq j\leq N;\label{(4)}
\\ && (E\psi_{1j},E\psi_{2j})=(\psi_{1j},\psi_{2j})U(u,\lambda_{j})^{-1},\enspace 1\leq j\leq N.\label{(5)}\end{eqnarray}
It is easy to see that 
\begin{equation}
M\frac {\delta \lambda_{j}}{\delta u}=\lambda_{j}J
\frac {\delta \lambda_{j}}{\delta u},\label{(8)}\end{equation}
where ${\delta \lambda_{j}}/{\delta u}$ is determined by 
\begin{equation}
\frac {\delta \lambda_{j}}{\delta u} = \left(\begin{array} {c} 
\frac {\delta\lambda_{j}}{\delta a} \vspace{2mm}\\
\frac { \delta\lambda_{j}}{\delta b}\end{array} \right)=\frac{\beta _j}{a}\left(
\begin{array} {c}
(\lambda_j-b)\phi_{2j}\psi_{1j}+\phi_{2j}\psi_{2j} \vspace{2mm}\\
a\phi_{2j}\psi_{1j}\end{array} \right),\ \beta _j=\textrm{const.}
\label{(7)} \end{equation}
Now the Bargmann constraint $G_{0}=\sum_{j=1}^{N}\beta _j^{-1}
\delta {\lambda_{j}}/ {\delta u}$ leads to a B\"acklund transformation
\begin{equation}
a=<A P_2,P_1>+<P_2,Q_2>-<P_2,Q_1>^2,\quad b=<P_2,Q_1>,\label{(9)}
\end{equation}
where $A={\rm diag}(\lambda_{1},\cdots,\lambda_{N}),\  P_{i}=(\phi_{i1},
\cdots,\phi_{iN})^{T}, \  Q_i=(\psi_{i1},\cdots,\psi_{iN})^{T}$, and 
$<\cdot,\cdot>$ is the standard inner product of 
$\mbox{\rm I \hspace{-0.9em} R}^{N}$, as defined before. 

Substituting (\ref{(9)}) into (\ref{(4)}) and (\ref{(5)}) yields a 
discrete Bargmann system
\begin{equation}\left\{\begin{array}{l}
EP_1=\left(<AP_2,Q_1>+<P_2,Q_2>-<P_2,Q_1>^2\right)P_2,
\vspace{2mm}\\
EP_2=-P_1-<P_2,Q_1>P_2+AP_2 ,%\label{(10)}
\vspace{2mm}\\
EQ_1=\displaystyle{\frac {Q_2-<P_2,Q_1>Q_1+AQ_1}
{<AP_2,Q_1>+<P_2,Q_2>-<P_2,Q_1>^2}\, ,}
\vspace{2mm}\\ 
EQ_2=-Q_1,\end{array} \right. \label{(10)}\end{equation}
which determines a symplectic mapping $H$:
\begin{equation}
(EP_1,EP_2,EQ_1,EQ_2)=H(P_1,P_2,Q_1,Q_2), \label{(11)}\end{equation}
since we have by a direct calculation
$$\sum^N_{j=1} d(E\phi^{(j)} )\wedge d(E\psi^{(j)})=\sum^N_{j=1}d\phi^{(j)}\wedge d\psi^{(j)}.$$
The generating function ${\mathcal F}_\lambda=\det V|_{u=f}$:
$${\mathcal F}_\lambda=-Q_\lambda(AP_1,Q_1)-Q_\lambda(P_1,Q_2)+<P_1,Q_1>Q_\lambda(P_2,Q_1)$$
$$+\left|\begin{array} {cc}
Q_\lambda(P_1,Q_1)&Q_\lambda(P_1,Q_2)\vspace{2mm}\\
Q_\lambda(P_2,Q_1)&Q_\lambda(P_2,Q_2)\end{array}
\right|=\sum_{m\geq 0}F_m\lambda^{-m-1}\label{(12)},$$
where 
$$ Q_{\lambda}(\xi,\eta)=\sum_{j=1}^{N}\frac{\xi_j\eta_j}{\lambda-\lambda_j}=\sum_{m\geq 0}<A^{m}\xi,\eta>\lambda^{-m-1},$$
generates a hierarchy of invariants of (\ref{(10)}):
$$F_{0}=<A P_1,Q_1>-<P_1,Q_2>+<P_1,Q_1><P_2,Q_1>,\label{(13)}$$
$$
F_{m}=-<A ^{m+1}P_1,
Q_1>-<A^mP_1,Q_2>+<P_1,Q_1>< A^m P_2,Q_1>$$
$$
+\sum_{i=1}^{m}\left|\begin{array} {cc}
<A^{i-1}P_1,Q_1>& <A^{m-i}P_2,Q_1>\vspace{2mm}\\
 <A^{i-1}P_1,Q_2>&<A^{m-i}P_2,Q_2>\end{array} \right|,\ m\ge 1.\label{(14)}$$
A direct computation can show the involutivity
$$\{ F_{m},\bar{F}_{l}\}=0, \enspace m,l\geq 0, \label{(16)}$$
where the variants $\bar {F}_j=\phi_{1j}\psi_{1j}+\phi_{2j}\psi_{2j}$, $1\le j\le N$,
defined as before.
Now we can easily see that the symplectic mapping (\ref{(10)}) is Liouville integrable 
\cite{BruschiRST-PD1991}.

Introduce an initial-value problem
\begin{equation}
\left .
\begin{array}{l}
 \displaystyle{
P_{it}=\frac{\partial F_0}{\partial Q_{i}}, \enspace  
Q_{it}=-\frac{\partial F_0}{\partial P_i},} %\label{(17a)}
%\vspace{2mm} \\
 \enspace \,   (P_i(t),Q_i(t))|_{t=0}=(P_{i0},Q_{i0}) ,\enspace i=1,2,% \label{(17b)}
\label{(17)}\end{array}\right. 
\end{equation}
where $P_{i0}$ and $Q_{i0},\ i=1,2,$ are arbitrary constant vectors.
Let $(P_i(t),Q_i(t)), \ i=1, 2,$ be a solution to
 the initial-value problems (\ref{(17)}), and further define 
\begin{equation}
(P_1(n,t),P_2(n,t),Q_1(n,t),Q_2(n,t))=H^n(P_1(t),P_2(t),Q_1(t),Q_2(t)).
\label{(18)}\end{equation}
Then $a(n,t)$ and $b(n,t)$
determined by the B\"acklund transformation (\ref{(9)})
solves the Toda lattice (\ref{(20)}).

\end{example}

\begin{example}
Let us now consider the Langmuir lattice \cite{ZakharovMNP-1980}: 
\begin{equation}
a_t(n,t)=a(n,t)(a(n+1,t)-a(n-1,t)),\label{Langmuir}\end{equation}
which associates with a reduction of the discrete spectral problem (\ref{(1)}):
\begin{equation}
E{\phi}=U(a,\lambda){\phi} ,\enspace  U(a,\lambda)=\left(\begin{array} {cc} 
0&a\vspace{2mm}\\  -1&\lambda\end{array} \right)
,\enspace \phi =\left(\begin{array} {c}\phi_1
\vspace{2mm}\\  \phi _2\end{array}\right)  .\label{(21)}
\end{equation}
Assume that its discrete zero-curvature equation 
\begin{equation} (EV)U-UV=0,\enspace V=(V_{ij})_{2\times 2}\label{(22)}\end{equation}
has a solution 
$$ V_{11}=aB-\lambda^2C,\quad V_{12}=\lambda E^{-1}aC,\quad V_{21}=-\lambda C,\quad V_{22}=E^{-1}aB , $$
where 
$$B=\sum_{i\geq 0}B_{i}\lambda^{-2i},\quad C=\sum_{i\geq 0}C_{i}\lambda^{-2i}.$$
Then upon choosing $B_0=C_0=1$, we can easily find that  
(\ref{(22)}) is equivalent to 
\begin{equation}
MB_{n-1}=JB_{n}, %\enspace B_0=1,\enspace j\geq 1\label{(23)}$$
\ C_n=(\Delta^*a-a\Delta )^{-1}\Delta^*aB_n,\ n\geq 1,
\end{equation} 
where two skew-symmetric operators $J$ and $M$ read as
\begin{equation}
M =a(\Delta-\Delta^*)a,\quad J=a\Delta(\Delta^*a-a\Delta)^{-1}\Delta^*a.\end{equation}
%and
%$$C_j=(a\Delta^*-a\Delta)^{-1}\Delta^*aB_j,\quad C_0=1, \quad j\geq 1.\label{(24)}$$ 
Let $\lambda_1,\cdots,\lambda_N$ be distinct eigenvalues, then 
we have
\begin{equation}
\left\{\begin{array}{l}
(E\phi_{1j},E\phi_{2j})=(\phi_{1j},\phi_{2j})U(a,\lambda_{j})^{T},\
 1\le j\le N,\vspace{2mm} \\
(E\psi_{1j},E\psi_{2j})=(\psi_{1j},\psi_{2j})U(a,\lambda_{j})^{-1},\ 
 1\leq j\leq N,\end{array}\right. 
\label{(25)}
\end{equation}
and
\begin{equation}
M\frac {\delta \lambda_{j}}{\delta a}=\lambda^2_{j}J
\frac {\delta \lambda_{j}}{\delta a},% \label{(26)}
\ \textrm{where}\  
%\begin{equation}
\frac {\delta \lambda_{j}}{\delta a}=\frac {\beta _j} a 
(\lambda_j\phi _{2j}\psi_{1j}+\phi_{2j}\psi_{2j} ),\ 
\beta _j=\textrm{const.}\label{(27)}\end{equation}
Now similarly, the Bargmann constraint $G_{0}=B_0=\sum_{j=1}^{N} {\beta _j}^{-1}
 {\delta   \lambda_{j}}/{\delta a}$ leads to 
a B\"acklund transformation 
\begin{equation}
a=<A P_2,Q_1>+<P_2,Q_2>.\label{(28)}
\end{equation}

Substituting (\ref{(28)}) into (\ref{(25)}), we obtain another 
discrete Bargmann system
 \begin{equation}
\left \{ \begin{array}{l}
E P_1=\left( <A P_2,Q_1>+<P_2,Q_2>\right)P_2,\ 
EP_2=-P_1+AP_2, %\label{(29)}$$
\vspace{2mm}\\
EQ_1=\displaystyle{\frac{Q_2+AQ_1}{<AP_2,Q_1>+<P_2,Q_2>}},\ EQ_2=-Q_1,\end{array}
\right. \label{(29)} 
\end{equation}
which determines a symplectic mapping $H$:
\begin{equation}
(EP_1,EP_2,EQ_1,EQ_2)=H(P_1,P_2,Q_1,Q_2). \label{(30)}
\end{equation}
This symplectic mapping
is Liouville integrable \cite{BruschiRST-PD1991}, since we have 
involutive invariants: $\bar{F}_j=\phi_{1j}\psi_{1j}+\phi_{2j}\psi_{2j},\ 
1\le j\le N,$ and $F_m$, $m\ge 0$, defined by 
\begin{equation}
\begin{array}{l}
{ F}_{0}=-<A^2P_1,Q_1>-<AP_1,Q_2>+<P_1,Q_1>(<AP_2,Q_1>+<P_2,Q_2>),%\label{(31)}
\vspace{2mm}\\
{F}_{m}=-<A^{2m+2}P_1,Q_1>-<A^{2m+1}P_1,Q_2>+<P_1,Q_1><A^{2m+1}P_2,Q_1>
%\nonumber \\ && \qquad
\vspace{2mm}\\
 +<P_2,Q_2><A^{2m}P_1,Q_1>
+\displaystyle{\sum_{i=1}^{m}}\left|\begin{array} {cc}
<A^{2i-2}P_1,Q_1>& <A^{2m-2i+1}P_2,Q_1>\vspace{2mm}\\
 <A^{2i-1}P_1,Q_2>&<A^{2m-2i+2}P_2,Q_2>\end{array} \right| .
\end{array} \nonumber \end{equation} 

Let $(P_i(t),Q_i(t)), \ i=1, 2,$ be a solution to an initial-value problem:
\begin{equation}\left.\begin{array}{l}
P_{it}=\displaystyle{
\frac{\partial { F}_0}{\partial Q_i}, \enspace  Q_{it}
=-\frac{\partial { F}_0}{\partial P_i}, }
%\label{(32a)}$$
%\vspace{2mm}\\
\enspace \, (P_i(t),Q_i(t))|_{t=0}=(P_{i0},Q_{i0}) ,\enspace 
i=1,2,
\end{array}\right.
%\label{(32b)}$$
\end{equation}
where $P_{i0}$ and $Q_{i0},\ i=1,2,$ are arbitrary, and similarly define 
\begin{equation}
(P_1(n,t),P_2(n,t),Q_1(n,t),Q_2(n,t))=H^n(P_1(t),P_2(t),Q_1(t),Q_2(t)).\end{equation}
Then $a(n,t)$ determined by the B\"acklund transformation
(\ref{(28)}): 
$$a(n,t)=<AP_2(n,t),Q_1(n,t)>+<P_2(n,t),Q_2(n,t)>$$
provides a solution to the Langmuir lattice (\ref{Langmuir}).

\end{example}

\section{Concluding remarks}

It has been shown that solving symmetry constraints for $u$ 
can give rise to B\"acklund transformations between soliton systems and lower-dimensional 
Liouville integrable systems, which supplements the study of binary nonlinearization of Lax pairs 
\cite{MaS-PLA1994,MaFO-PA1996,Ma-JPSJ1995,MaF-book1996}.
Construction of solutions to soliton systems is split into finding  
solutions to the space and time parts of integrable 
constrained Lax pairs, which gives  
a way to separate variables for soliton systems
and exhibits integrability by quadratures for soliton systems.
Upon solving the Riemann-Jacobi inversion problems for constrained Lax pairs,  
the resulting B\"acklund transformations can generate finite-gap solutions to soliton
systems in terms of Riemann-theta functions.

We remark that all symmetry constraints defined by (\ref{sc}) 
can put Lax pairs into integrable symplectic mappings and/or integrable 
finite-dimensional Hamiltonian systems. 
The corresponding constrained Lax pairs may have some
specific properties, e.g., 
bi-Hamiltonian and quasi-bi-Hamiltonian structures.
Therefore, symmetry constraints are very powerful in constructing 
lower-dimensional integrable systems from Lax pairs of soliton systems.
Nevertheless, there exist symmetry constraints which do not force Lax pairs into integrable systems \cite{MaL-1999}, and the problem of integrability has not been solved for the time
parts of the original constrained Lax pairs, i.e., systems of partial differential equations
\begin{equation} \left\{\begin{array}{l}
\phi _{t_n}^{(j)}= {V} ^{(n)}(f,f_x,\cdots;\lambda _j)\phi ^{(j)}
,\ 1\le j\le N,\vspace{2mm}\\
 \psi ^{(j)}_{t_n}=- V^{(n)T}(f,f_x,\cdots;\lambda _j)\psi ^{(j)},\ 1\le j\le N;\end{array}
\right. \end{equation}
and systems of difference-differential equations
\begin{equation} \left \{ \begin{array}{l}
\phi _{t_n}^{(j)}= {V} ^{(n)}(f,Ef,E^{-1}f,\cdots;\lambda _j)\phi ^{(j)}
,\ 1\le j\le N,\vspace{2mm}\\
 \psi ^{(j)}_{t_n}=-V^{(n)T}(f,Ef,E^{-1}f,\cdots;\lambda _j)\psi ^{(j)},\ 
1\le j\le N.\end{array}\right.
\end{equation}
We are curious to known whether they are good candidates for 
integrable systems.
%The resulting B\"acklund transformations are still valid in 
%relating soliton systems to these systems of partial differential equations
%and difference-differential equations.

\bibliographystyle{amsalpha}

\end{document}